\documentclass[12pt]{article}
\usepackage{amsmath,amssymb,amsfonts,amsthm}
\usepackage{graphicx}
\usepackage{epsfig,multicol}

\newcommand{\be}[0]{\begin{equation}}
\newcommand{\ee}[0]{\end{equation}}
\newcommand{\ba}[0]{\begin{eqnarray}}
\newcommand{\ea}[0]{\end{eqnarray}}

\title{Meson spectroscopy in a confining theory via AdS/CFT}

 \author{ Majid Dehghani
\\
\footnotesize{Physics Group, Azad University of Marvdasht, Marvdasht, Fars, Iran}
\\ \footnotesize{e-mail:m.dehghani55@gmail.com  }
\\
Behrouz Mirza
\\
\footnotesize{Department of Physics, Isfahan University of Technology, Isfahan, 84156-83111, Iran }
\\ \footnotesize{e-mail: b.mirza@cc.iut.ac.ir  } }

\begin{document}

 \date{\today}
\maketitle



 \newcommand{\norm}[1]{\left\Vert#1\right\Vert}
 \newcommand{\abs}[1]{\left\vert#1\right\vert}
 \newcommand{\set}[1]{\left\{#1\right\}}
 \newcommand{\R}{\mathbb R}
 \newcommand{\I}{\mathbb{I}}
 \newcommand{\C}{\mathbb C}
 \newcommand{\eps}{\varepsilon}
 \newcommand{\To}{\longrightarrow}
 \newcommand{\BX}{\mathbf{B}(X)}
 \newcommand{\HH}{\mathfrak{H}}
 \newcommand{\D}{\mathcal{D}}
 \newcommand{\N}{\mathcal{N}}
 \newcommand{\la}{\lambda}
 \newcommand{\af}{a^{ }_F}
 \newcommand{\afd}{a^\dag_F}
 \newcommand{\afy}{a^{ }_{F^{-1}}}
 \newcommand{\afdy}{a^\dag_{F^{-1}}}
 \newcommand{\fn}{\phi^{ }_n}
 \newcommand{\HD}{\hat{\mathcal{H}}}
 \newcommand{\HDD}{\mathcal{H}}

\abstract{Instanton effects may have implications for hadronization of quark-gluon plasma as it cools. Here we study
dispersion relations of mesons in a strongly coupled plasma with an instanton background present. It will be shown that
at higher energies the instanton effect diminishes and some comments on the limiting velocity of mesons in the plasma.
the profile function of mesons on the gravity side is considered also because of its relevance to energy loss of quark in plasma. }

 \section{Introduction}\label{sec-intro}

 Quark-Gluon Plasma is a new state of matter created in high-energy collisions of heavy nuclei. It is the result of high energy-density limit in quantum chromodynamics (QCD) theory. The existence of QGP was predicted long ago at the time when QCD was formulated as a gauge theory of strong interactions \cite{collins1975superdense,shuryak1978theory}. Based on the asymptotic freedom of QCD at short distances it was argued that in a dense matter at high temperature the quarks and gluons are weakly interacting. After a long time, a strong experimental evidence for the existence of QGP at RIHC has been reported, and data from experiment showed that the system is strongly coupled and needs non-perturbative treatment. Theoretical investigations before the arrival of experimental data, were based on perturbative calculations. Limited tools are available in quantum field theory (QFT) for investigating strongly coupled theories (lattice gauge theory) and specially get into trouble at finite temperatures. So any new tool for investigating a strongly coupled field theory is valuable.

 One of the tools in investigating finite temperature QCD, coming from string theory is the AdS/CFT correspondence \cite{maldacena1999large}. It maps a strongly coupled large-N gauge theory to weakly coupled supergravity theory. The gravity side is the near horizon limit of $N$ D3 brane. The theory with fundamental matter present is obtained by introducing probe D7 branes in the theory \cite{karch2002adding}.

 On the other hand instanton liquid model (ILM) is successful in describing many aspects of hadron physics, but it is a computationally difficult procedure. Some times in hadron physics the instanton gas model (instantons without interaction), which is computationally much simpler, is used to extract some features of hadrons.  Lattice calculations show that instanton gas model is not a good approximation in the hadronic phase. It seems that the right place for the instanton gas model is in the finite temperature  phase of QCD.
  Instantons QCD  are responsible for chiral symmetry breaking and it is known that at finite temperature the chiral symmetry is restored. So one may conclude that the restoration of chiral symmetry at finite temperature is the result of instanton disappearance. But there is evidence from lattice simulations that instantons survive the  phase transition \cite{schaefer1998instantons}. The chiral symmetry restoration is the result of a change in instanton configuration, i.e. forming molecules of instanton/anti-instanton. These non-perturbative effects do not survive too far above the transition temperature ($T_C$).

 From gauge/gravity correspondence we know that the dual of Yang-Mills instantons are the D-instantons of gravity side. More precisely it is the correspondence between type-II B string theory in near-horizon D3+D(-1) geometry and N=4 SYM theory in a constant self dual gauge field background \cite{hong1999d3}. The self dual gauge field is identified with gluon condensation
 and as explained in \cite{gwak2012holographic}, from the gravity point of view is responsible for both chiral symmetry breaking and confinement. Using AdS/CFT correspondence, it is shown in \cite{hong1999d3} that the D-instanton background is a confining theory with linear quark-antiquark potential.
 Meson spectral functions in strongly coupled plasma was considered in \cite{casalderrey2010new, ejaz2008limiting, casalderrey2010cherenkov}. These works rely on the existence of two kinds of D7 brane embedding at finite temperature. One is the Minkowski embedding where brane stays out of the horizon and the other is black hole embedding where brane meets the horizon. Stable quasi-particles exist for the Minkowski embedding but not for the other one.

 They argued that although the $\mathcal{N}=2$ SYM theory is in a deconfined phase, heavy quarkonium mesons can exist for sufficiently low temperatures of Minkowski embedding near critical temperature ($T_c$). Then mesons in the plasma are gapped and have discrete spectrum.  This is exactly the region where instantons exist near and above the critical temperature, so they can affect the properties of  quarkonium mesons.
 The D7 brane embedding at finite temperature and a dilaton background present was considered in \cite{ghoroku2005flavor, gwak2012holographic}. In \cite{ghoroku2005flavor} the dependence of thermal quark mass and the effective quark-antiquark potential on instanton density  was considered. The issue of chiral symmetry breaking and phase transition temperature in the same background was studied in \cite{gwak2012holographic}.

  Here we will consider the effect of instantons on the spectral function of these heavy mesons. The organization of this paper is as follows: in section \textbf{\ref{geometry}} we will introduce the geometry of D3-D(-1) brane configuration. Using the introduce background geometry, in section  \textbf{\ref{meson spec}} the spectral function for vector and scalar  mesons is obtained numerically. we also show the meson profile functions for different instanton density and momentum numbers.


\section{ ِD3/D-Instanton Geometry}\label{geometry}


As explained in the introduction, the geometry concerned here is that of D3/D-instanton background at finite temperature \cite{ghoroku2005flavor}.
Here the near horizon limit of $D3-D(-1)$ background with Euclidean signature is considered. The D7 brane will be embedded in this background and the fluctuations of brane around embedding will be considered as mesons of theory.
Ten dimensional supergravity action with five-form and axion field in Einstein frame is given by \cite{gibbons1996instantons,gwak2012holographic} :

\begin{equation}\label{action10D}
  S=\frac{1}{\kappa}\int d^{10}x \sqrt{g} \left(R-\frac{1}{2}(\partial\Phi)^2 +\frac{1}{2}(\partial\chi)^2-\frac{1}{6}F^2_{(5)} \right)
\end{equation}

where $\Phi$ and $\chi$ are dilaton and axion fields respectively. If we set $\chi=-e^{-\Phi}+\chi_0$, in the above action, then dilaton and axion terms cancel. Then the solution with metric and five-form field in string frame is: \cite{mateos2007thermodynamics}:

\begin{eqnarray}\label{finit temp}
  ds^2=e^{\Phi/2}\left[\frac{u^2}{L^2}(-f(u)dt^2+dx^2)+\frac{L^2}{u^2}(\frac{du^2}{f(u)}+u^2d\Omega^2_5)\right]    \\
   e^\Phi=1+\frac{q}{r^4_0}log\frac{1}{f(u)^2},\;\;\;\;\;\ \chi=-e^{-\Phi}+\chi_0,\;\;\;\;\;\ \nonumber  \\
 f(u)=\sqrt{1-\left(\frac{u_0}{u}\right)^4},\;\;\;\;\;\;\;\;\;\;\;\;\;\;\;\;\;\;\;\;\;\;\
\end{eqnarray}
where the horizon lies at $u_0$,  and $d\Omega_5^2$ is the unit five-sphere $S_5$ metric.

For the D7-brane embedding considerations, it is useful to perform the following coordinate change:

\begin{equation}\label{coord chang}
  u_0^2 \rho^2=u^2+\sqrt{u^4-u_0^4}
\end{equation}
where the black hole horizon corresponds to $\rho=1$ and the Minkowski boundary to $\rho\rightarrow \infty$.
Then the 10 dimensional metric becomes:

\begin{equation}\label{BHmetric}
  ds^2= e^{\Phi/2}  \frac{\rho^2}{2L^2} \left[ -\frac{f^2}{\tilde{f}^2}dt^2+\tilde{f}dx^2_i]+ \frac{L^2}{\rho^2}[dr^2+r^2d\Omega^2_3+dR^2+R^2d\vartheta  \right]
\end{equation}
where:

\begin{equation}\label{BH met2}
  L^4=\pi g_s N_c l^2_s,\;\;\;\;\;\;\;\  \rho^2=R^2+r^2,\;\;\;\;\;\;\;\  f=1-\frac{1}{\rho^4},\;\;\;\;\;\;\;\ \tilde{f}=1+\frac{1}{\rho^4}
\end{equation}
Here the directions spanned by D7 brane are $(t,\vec{x},r)$, wraps $S^3$ and perpendicular to $R$ and $\vartheta$ directions. Using the rotational symmetry in the transverse coordinates of D7 brane, we can set $\vartheta=0$. The induced metric on D7 brane follows \ref{BHmetric} as:

\begin{equation}\label{D7 induced}
  ds^2_{D7}= e^{\Phi/2}  \frac{\rho^2}{2L^2} \left[ -\frac{f^2}{\tilde{f}^2}dt^2+\tilde{f}dx^2_i]+ \frac{L^2}{\rho^2}[(1+\dot{R}^2)dr^2+r^2d\Omega^2_3  \right]
\end{equation}
where $\dot{R}=dR/dr$ and $R$ is the profile of D7 brane embedding.

\section {Meson Dispersion Relations in N=2 SYM Plasma }\label{meson spec}

Similar to the case of D7 brane at finite temperature without instantons, there are two types of embeddings. One is the low temperature Minkowski embedding where D7 brane sit outside the black hole horizon and stable mesons with mass gap and discrete spectrum  exist. The other one is high temperature black hole embedding where D7 brane meets the black hole horizon where meson spectrum is gapless and continuous.
Here we consider mesons in the Minkowski embedding of the confining theory, explained in previous section. Here, the repulsion of D7-brane due to instanton background prevents the D7 brane to be pulled towards the horizon and results in stable mesons for a wider range above the critical temperature ($T_c$). The scalar and vector meson
on the gauge theory side corresponds to the regular and normalizable modes of scalar and vector fields on the D7-brane \cite{casalderrey2010cherenkov}. The world volume gauge fields of the embedded brane correspond to vector mesons and the geometric fluctuation of brane around the embedding correspond to scalar mesons.

\subsection{vector mesons}

The fluctuations of D7 brane world volume gauge fields give are dual to vector mesons of boundary. These world volume gage fields must be regular and normalizable. With the use of quadratic order gauge field action the linearized equation of motion
for  vector meson modes will be obtained. Here we will analyse the transverse vector mesons only.
The DBI action of D7 brane quadratic in gauge field reads \cite{mateos2007thermodynamics, myers2007holographic}:

\begin{equation}\label{DBI gauge}
  S_{D7}=-\frac{(2\pi l_s)}{4} T_{D7}N_f \int d^8\sigma \sqrt{-g} g^{mp}g^{nq}F_{mn}F_{pq},
\end{equation}
where $g_{mn}$ is the induced metric on D7 brane \ref{D7 induced}, $T_{D7}=2\pi/(2\pi)^8g_s$ and $F_{mn}=\partial_m A_n-\partial_n A_m$.
The above action leads to the following equation of motion:

\begin{equation}\label{gaugeEOM}
  \partial_m(\sqrt{-g}F^{mn})=0
\end{equation}
Here the singlet modes are concerned, i.e. modes independent of $S^3$-components and also the gauge choice for radial component of gauge field $A_r=0$. So:

\begin{equation}\label{gauge modes}
  A_{\mu}= A_{\mu}(x^{\mu},r), \;\;\;\;\;\;\     A_r=0, \;\;\;\;\;\;\    A_{\Omega_3}=0,
\end{equation}
The remaining four components are plane waves in Minkowski directions with Fourier components defined through:

\begin{equation}\label{Foureir gauge}
  A_{\mu}(t,x,r)=\int \frac{d\omega d^3k}{(2\pi)^4}  A_{\mu}(\omega,k,r) e^{-i\omega+ik.x},
\end{equation}
where $k$ and $\omega$ are three momentum and energy of meson, respectively. As in \cite{casalderrey2010cherenkov}, the momentum is pointed along $x^1$ direction so the transverse modes are $A_2$ and$ A_3$ and their equation of motion decouple from each other and from those for longitudinal modes $A_0,A_1$.

First we will consider the transverse modes $A_i$ with $i=1,2$. The equation of motion for transverse modes takes the form:

\begin{equation}\label{transv EOM}
  \partial_r(e^{3\Phi}\sqrt{-g}g^{rr}g^{33}\partial_r A_i )- e^{3\Phi}\sqrt{-g}g^{33}(g^{00}\omega^2+ g^{11}k^2 )=0
\end{equation}
Using Eq.(\ref{D7 induced}) results in:

\begin{equation}\label{transv EOM2}
   \partial_r \left(e^{3\Phi} \frac{fr^3}{2\sqrt{1+\dot{R}^2}} \partial_r A_i \right) + e^{3\Phi} \sqrt{1+\dot{R}^2}\frac{r^3}{\rho^4} \left(\frac{\omega^2 \tilde{f}}{f}- \frac{k^2 f}{\tilde{f}}   \right) A_i=0
\end{equation}
Now we expand transverse modes in terms of regular and normalizable modes:

\begin{equation}\label{expansion gauge}
  A_i(\omega,k,r)=\sum_n A_n(\omega,k)\xi_n(k,r)
\end{equation}
These normalizable modes with eigenvalues $\omega=\omega_n(k)$ obey the following equation:

\begin{eqnarray}\label{transv EOM2}
   \partial_r \left(e^{3\Phi} \frac{fr^3}{2\sqrt{1+\dot{R}^2}} \partial_r \xi_n(k,r) \right) + e^{3\Phi} \sqrt{1+\dot{R}^2}\frac{r^3 f}{\rho^4 \tilde{f}}k^2 \xi_n(k,r)                                \nonumber  \\
    = e^{3\Phi} \sqrt{1+\dot{R}^2}\frac{r^3 \tilde{f}}{\rho^4 f} \omega_n(k)^2 \xi_n(k,r)
\end{eqnarray}

As explained in \cite{casalderrey2010cherenkov}, regularity at $r=0$ and normalizability at $r=\infty$ results in the discreetness of the spectrum, where each of these fields is dual to a vector meson on the gauge theory side. $n$ is the internal quantum number of these 'Kaluza-Klein-reduced' tower of  modes.

\begin{figure}
\begin{center}$
\begin{array}{cc}
\includegraphics[width=7cm]{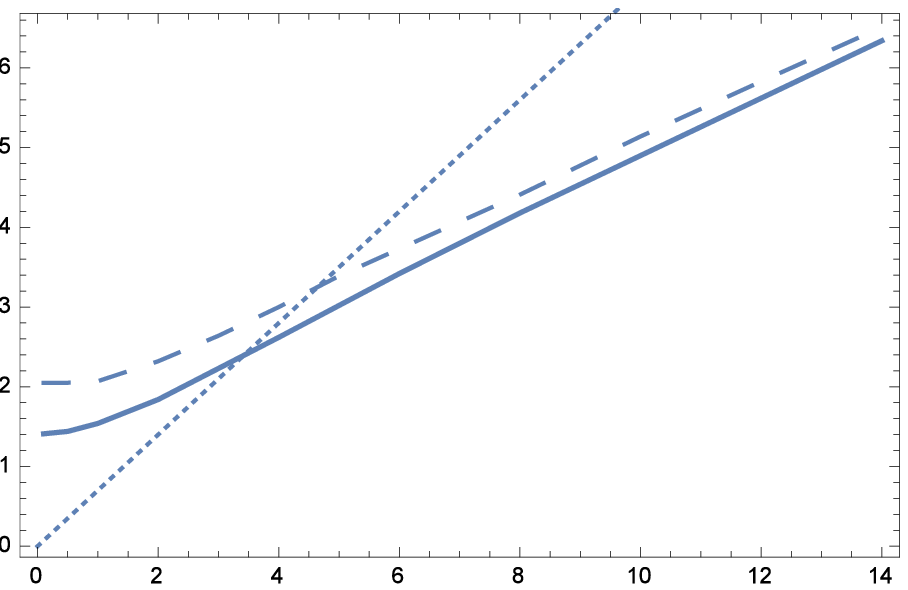} &
\includegraphics[width=7cm]{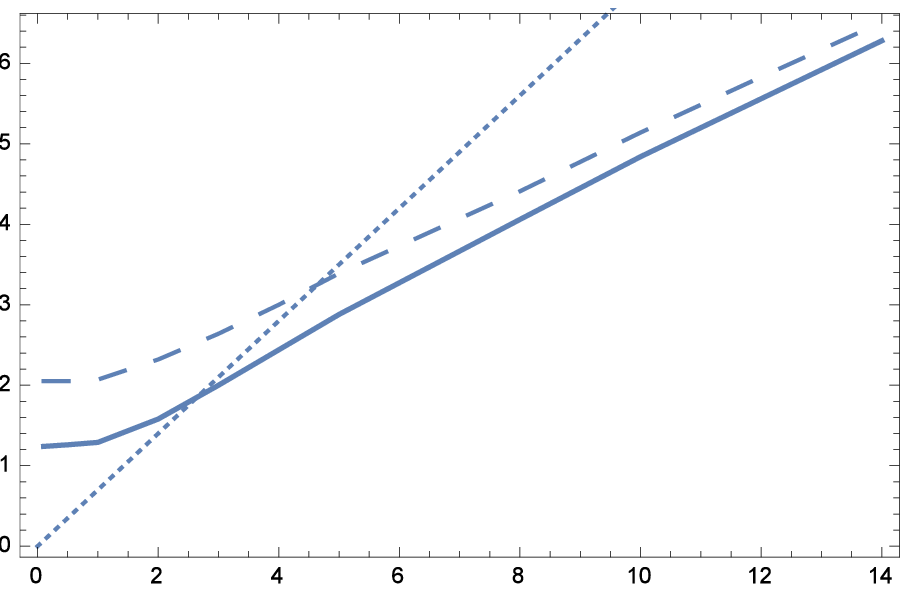}
\end{array}$
\end{center}
\caption{\footnotesize Vector meson dispersion relation at finite temperature with (thick line) and  without (dashed) instanton background, left: for instanton density $q=0.5$, right: for instanton density $q=1$.   }\label{vec fig}
\end{figure}
We have plotted the dispersion relation  $\omega_n(k)$ numerically in Figure [\ref{vec fig}].
We see from the figures that the instanton background changes the dispersion relations at low energies and as the energy goes to infinity, it approaches the value without instanton background. This means that the limiting velocity of vector meson do not change in the instanton background. This conclusion can be inferred from another point of view. The limiting velocity of the meson can be obtained from:

\begin{equation}\label{limiting veloc}
  \upsilon_{lim}=\sqrt{-\frac{g_{00}}{g_{11}}}
\end{equation}
and from the induced metric on D7 brane (\ref{D7 induced}), we can see that the dilaton factor is eliminated from the numerator and denominator of above equation, so the same limiting velocity in cases with and without dilaton background.
On the gauge theory side the $n$ in expansion (\ref{expansion gauge}) corresponds to different internal quantum numbers of mesons. On the gravity side, each value of $n$ gives a $k$-dependent radial wave function $\xi_n (k,r)$. As explained in \cite{casalderrey2010cherenkov}, this r-dependent profile determines the strength with which these modes couple to a quark.
As we will see, the instanton background changes the profile and therefore the coupling to quark. Then it can be concluded that the instanton background affects the energy loss of a quark moving through quark-gluon plasma.

In figure \ref{profile funs} we have compared different profile functions with and without instanton background. Figures shows that the presence of instantons moves the profile function towards higher $r$ values. To interpret this result we mention the result from \cite{gwak2012holographic} that the  Minkowski embedding of D7 brane with a D-instanton background experiences a repulsive force. So this repulsion prevent the wavefunction of mesons to become concentrated around the bottom of the D7 brane, $r\simeq 0$.

\begin{figure}
\begin{center}$
\begin{array}{cc}
\includegraphics[width=7cm]{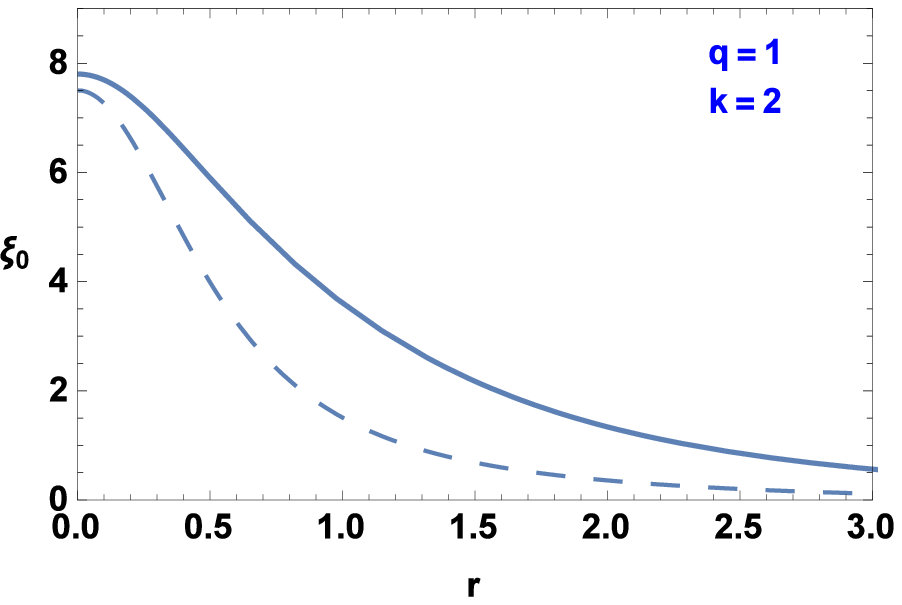} &
\includegraphics[width=7cm]{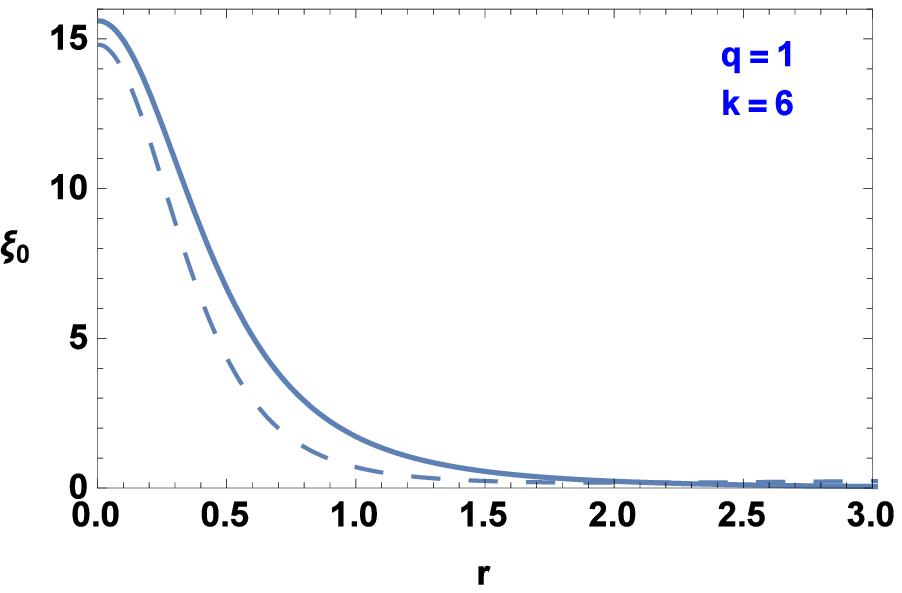} \\
\includegraphics[width=7cm]{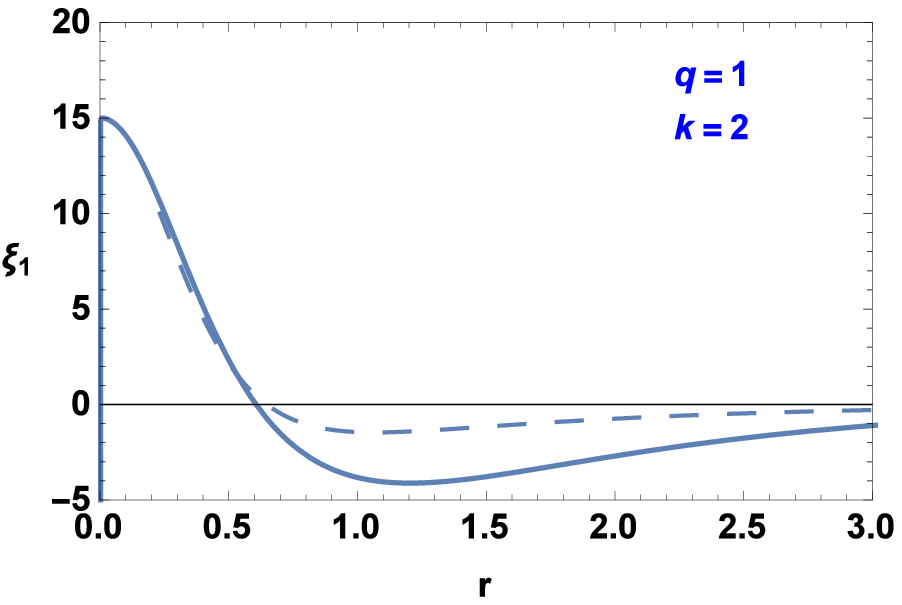} &
\includegraphics[width=7cm]{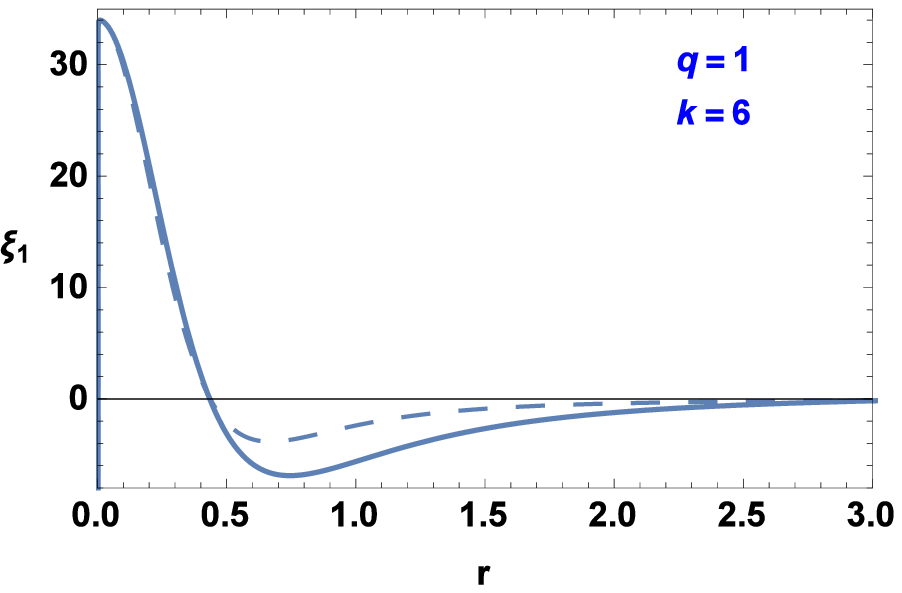}
\end{array}$
\end{center}
\caption{\footnotesize Vector meson profile functions. k is three-momentum and q instanton density number.Instanton background causes the profile functions to move towards higher $r$ values.}\label{profile funs}
\end{figure}

\subsection{scalar mesons}

Scalar mesons correspond to geometric fluctuations of D7 brane about the embedding in the background concerned here i.e. D3-D(-1) background.  String endpoint pulls the brane and excites the fluctuations on it. As explained in \cite{casalderrey2010cherenkov, ejaz2008limiting} the string end point is orthogonal to the D7 bane and a geometric way to compute the small fluctuations is to introduce local perpendicular unit vectors. The details of the formalism taken from \cite{casalderrey2010cherenkov, ejaz2008limiting} and used here can be find in appendix \ref{appena}.

The final form of scalar (and pseudoscalar) equation of motion becomes:

\begin{eqnarray}\label{scalar EOM app}
  \partial_r \left( e^{5\Phi/2} \frac{f \tilde{f}r^3 \rho^2 }{\sqrt{1+\dot{R}^2}}  \partial_r X^A  \right)+ e^{5\Phi/2} f \tilde{f}r^3 \sqrt{1+\dot{R}^2} \left( \frac{2\tilde{f}}{\rho^2 f^2}\omega^2-   \frac{2}{\rho^2 \tilde{f}}k^2 \right) X^A \nonumber \\
   - e^{2\Phi} f \tilde{f}r^3 \sqrt{1+\dot{R}^2} m^2 X^A =0
\end{eqnarray}

Then expanding $X^A$ in terms of normalizable functions $\phi_n^A(k,r)$, as:

\begin{equation}\label{scalar expans}
  X^A(\omega,q,r)=\sum_n X_n^A(\omega,k)\phi_n^A(k,r)
\end{equation}

\begin{figure}
\begin{center}$
\begin{array}{cc}
\includegraphics[width=7cm]{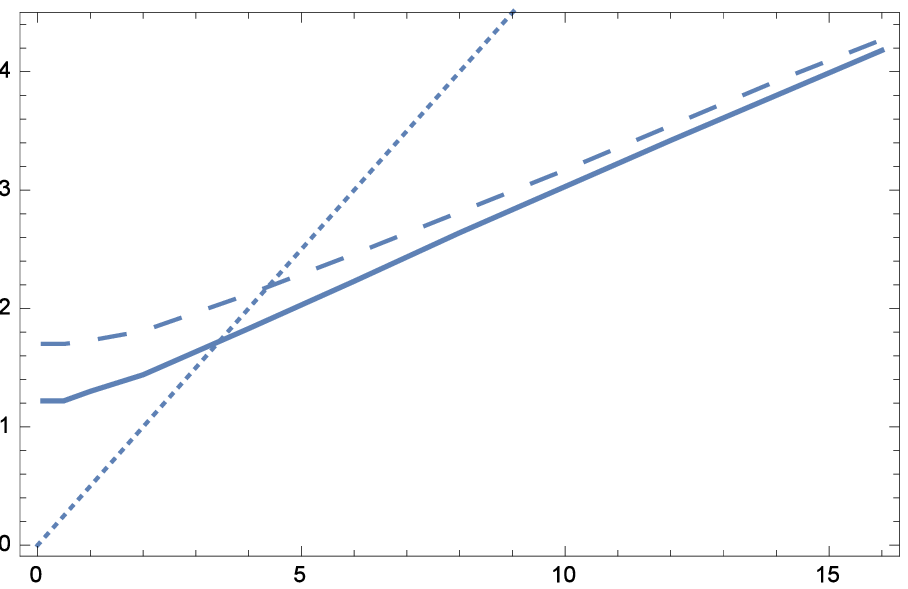} &
\includegraphics[width=7cm]{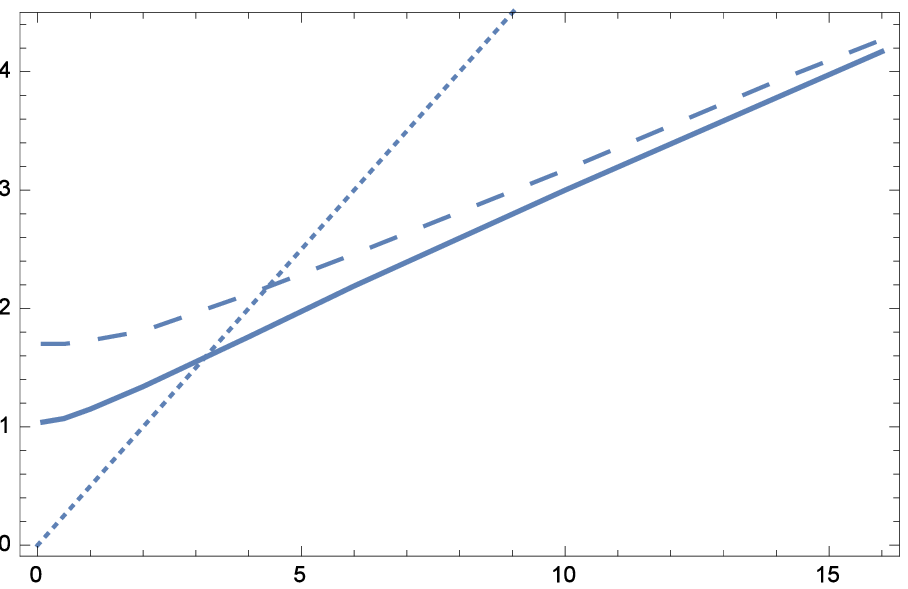}
\end{array}$
\end{center}
\caption{\footnotesize Scalar meson dispersion relations at finite temperature with (thick line)  and  without (dashed) instanton  background. left: for instanton density $q=0.5$, right: for instanton density $q=1$.  }\label{scalar fig}
\end{figure}

Then we obtain the dispersion relation by inserting (\ref{scalar expans}) in (\ref{scalar EOM}). Solving the dispersion relation numerically results in figure\ref{scalar fig} , which compares dispersion relations with and without instanton background.

Again solving the eigenvalue equation for each $n$ in Eq.(\ref{scalar expans}), we obtain $k$-dependent profile function $\phi_n^A(k,r)$ for scalar and pseudoscalar mesons. The results for the cases with and without instanton background are shown in figure [....] where here also the instanton effect on profile function is to repel it from the bottom of the brane $r\simeq 0$.

 \begin{figure}
\begin{center}$
\begin{array}{cc}
\includegraphics[width=7cm]{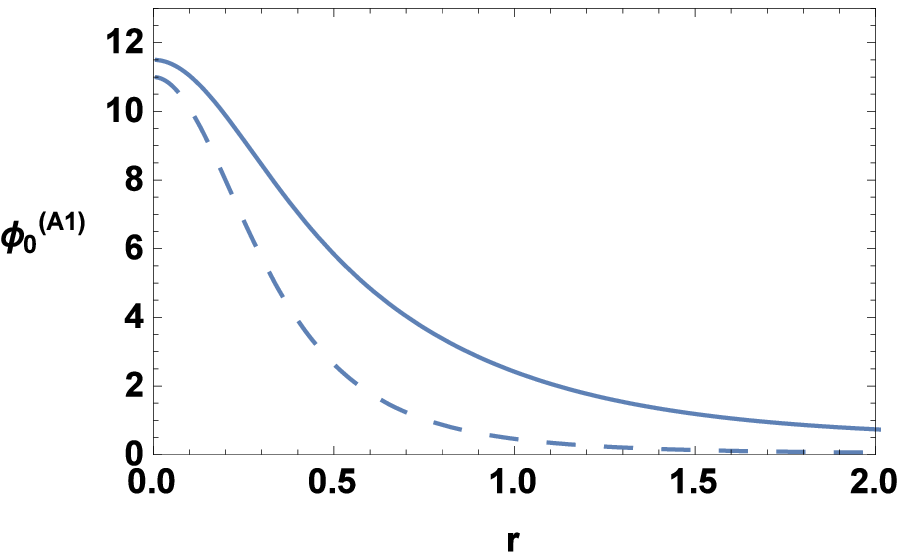} &
\includegraphics[width=7cm]{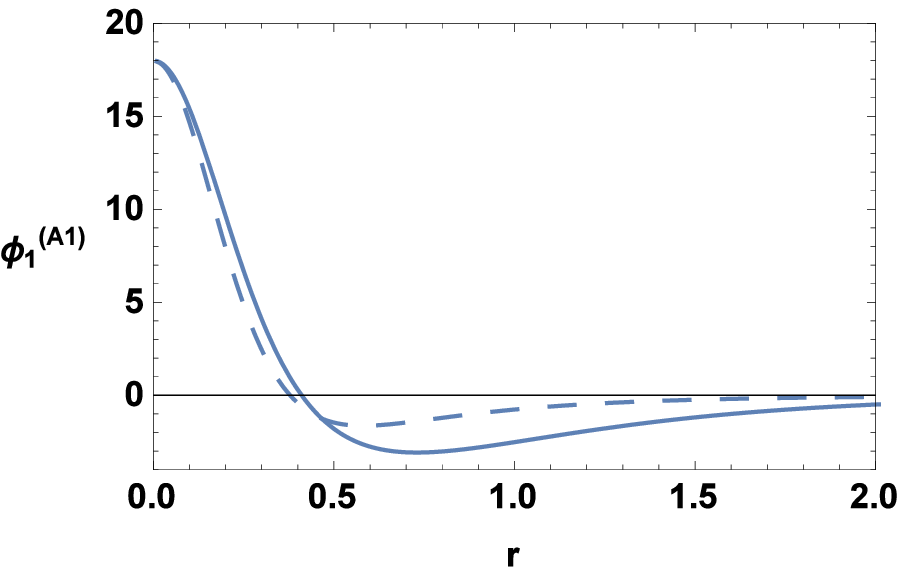} \\
\includegraphics[width=7cm]{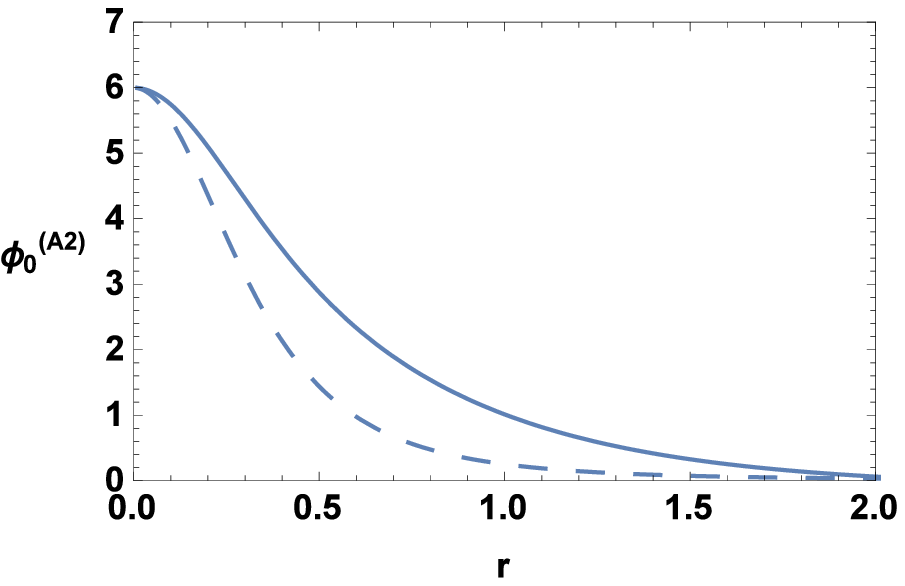} &
\includegraphics[width=7cm]{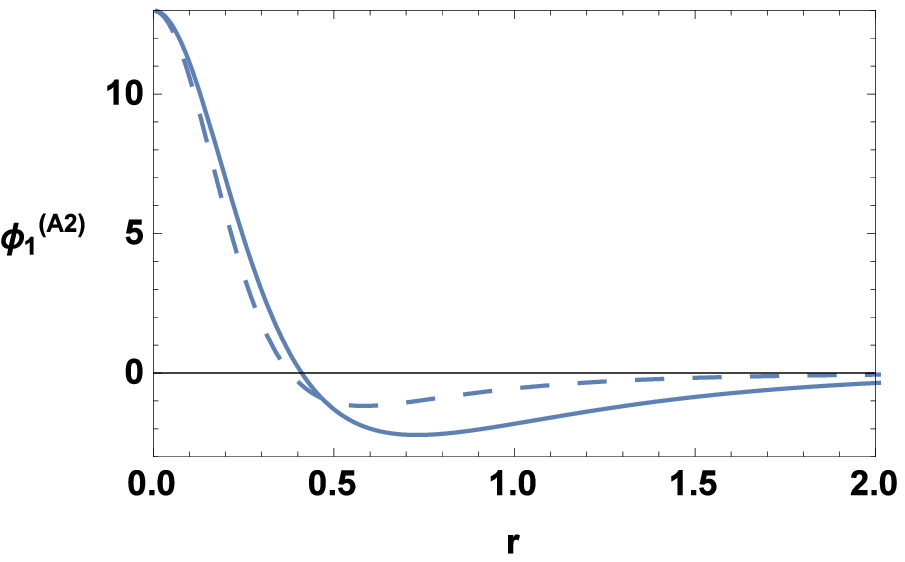}
\end{array}$
\end{center}
\caption{\footnotesize Scalar and pseudoscalar meson profile functions. k is three-momentum and q instanton density number.Top :Scalar meson, Bottom:pseudoscalar meson.  k is three-momentum and q instanton density number. Instanton background causes the profile functions to move towards higher $r$ values.  }\label{profile funs}
\end{figure}

\section {SUMMARY}\label{conclu}
We studied the quarkonium meson dispersion relations at finite temperature in the presence of an instanton background.
We showed how dispersion relations change specially at low energies. It is known that instanton background ripples the embedded D7 brane from the horizon and by calculating the profile functions as a function of fifth dimension at gravity side, we saw that results are consistent with this picture. The profile functions of mesons determine the coupling of quark in plasma with mesons, so the amount of energy loss. As a future work the implications of instanton background on the quark energy loss can be studied following our calculations.

\appendix
\numberwithin{equation}{section}
\section {Appendix}\label{appena}

 Here we present the scalar meson fluctuation around the embedding of D7 brane in the D3-D(-1) background following \cite{casalderrey2010cherenkov}. Because boundary conditions at string endpoint imply that strings must end orthogonal to the brane, for a clear geometric interpretation it is better to introduce spacetime coordinates that locally parametrize the directions orthogonal to the brane. The unite vectors orthogonal to the brane at each point of the brane, are:

 \begin{equation}\label{unit vectors}
   V_1 \propto \frac{\partial}{\partial R}-\dot{R}\frac{\partial}{\partial r},  \;\;\;\;\;\;\;\;\;\  V_2 \propto \frac{\partial}{\partial \vartheta}
 \end{equation}
Any vector orthogonal to the brane takes the form $U=X^A V_A$, so the $X^A$ are the coordinates orthogonal to the brane since by construction we have $\partial/\partial X^A=V_A $.
Then the fiducial embedding of brane is given by $X^A=0$, so that the $X^A$ field parametrize  fluctuations around it. Then the D7-brane action to quadratic order becomes:

\begin{equation}\label{scalar action}
  S_{scalar}=-T_{D7}L^8 \int d^8x \sqrt{-g} \left[\frac{1}{2}g^{ab}\partial_a X^A \partial_b X^B G_{AB}+\frac{1}{2} m_{AB}^2(x)X^A X^B \right]
\end{equation}
where $g$ is induced metric (\ref{D7 induced}) and $X^A$ independent. The diagonal mass matrix $m_{AB}^2$
is given in terms of the following geometric quantities:

\begin{align}\label{Ricci curv}
 m_{11}^2 &= R_{11}+R_{2112}+2R_{22}+ R^{(8)}-R,        \\  \nonumber
 m_{11}^2 &= -R_{22}+R_{2112}
\end{align}
where

\begin{align}\label{Ricci curv}
 R_{2112} &= V_2^M V_1^N V_1^P V_2^Q R_{MNPQ}      \\
  R_{11}&=  V_1^M V_1^N R_{MN}                     \\
  R_{22}&=  V_2^M V_2^N R_{MN}
\end{align}
$ R_{MNPQ}$ and $R_{MN}$ are Reimann and Ricci tensors of ten-dimensional spacetime metric $G$, respectively, and $R$ the corresponding Ricci scalar. $ R^{(8)}$ is the Ricci scalar of the eight-dimensional induced metric on D brane $g$. Action (\ref{scalar action}) with $G_{AB}=\delta_{AB}$ leads to the following equation of motion for scalar meson:

\begin{equation}\label{scalar eq}
  \partial_{a}(\sqrt{-g}\partial^a X^A)- \sqrt{-g}m^2 X^2 =0
\end{equation}
 $m= m_{11}$ or $ m_{22}$. For the case of $X^A$ independent of the $S^3$ and using its Fourier transformed
  component $X^A(\omega,k,r)$, then the above equation becomes:

  \begin{equation}\label{scalar eq Fouri}
     \partial_{r}(e^{5\Phi/2} \sqrt{-g} g^{rr} \partial^r X^A)- e^{5\Phi/2} \sqrt{-g}(g^{00}\omega^2 +g^{11}q^2)- e^{2\Phi}m^2 X^A=0
  \end{equation}
Now by substituting the induced metric (\ref{D7 induced}) the equation of motion is:

\begin{eqnarray}\label{scalar EOM app}
  \partial_r \left( e^{5\Phi/2} \frac{f \tilde{f}r^3 \rho^2 }{\sqrt{1+\dot{R}^2}}  \partial_r X^A  \right)+ e^{5\Phi/2} f \tilde{f}r^3 \sqrt{1+\dot{R}^2} \left( \frac{2\tilde{f}}{\rho^2 f^2}\omega^2-   \frac{2}{\rho^2 \tilde{f}}k^2 \right) X^A \nonumber \\
   - e^{2\Phi} f \tilde{f}r^3 \sqrt{1+\dot{R}^2} m^2 X^A =0
\end{eqnarray}



\vspace {2 cm}




\begin{thebibliography}{9}



\bibitem{maldacena1999large} J. Maldacena.   , \textit{The large-N limit of superconformal field theories and supergravity,   }     Int. j. theor. phys. {\bf 38 },1113 (1999).

\bibitem{hong1999d3} H. Liu, A. A. Tseytlin, \textit{D3-brane-D-instanton configuration and N= 4 super YM theory in constant self-dual background,}     Nuc. Phys. B. {\bf 553 },231 (1999)[arXiv:hep-ph/9903091];

 \bibitem{shuryak1978theory} E. V. Shuryak. ,  \textit{ Theory of hadron plasma,  }    Sov. Phys.-JETP  {\bf 47 },pg (1978).

 \bibitem{collins1975superdense} J.C. Collins, Perry,J. Malcolm, \textit{  Superdense matter: neutrons or asymptotically free quarks?, } Phys. Rev. Lett. {\bf 34}, 1353 (1975).

 \bibitem{schaefer1998instantons}T. Schaefer and E. V. Shuryak.,  \textit{ Instantons in QCD,  }     Rev. Mod. Phys. {\bf 70 }, 323 (1998)[arXiv:hep-ph/9610451];

 \bibitem{karch2002adding} A. Karch and E. Katz,  \textit{ Adding flavor to AdS/CFT  }   JHEP  {\bf 2002 },043 (2002)[arXiv:hep-th//0205236];


\bibitem{ghoroku2005flavor} K. Ghoroku, T. Sakaguchi, N. Uekusa, and M. Yahiro,  \textit{ Flavor quark at high temperature from a holographic model,  }    Phys. Rev. D  {\bf 71 }, 106002 (2005)[arXiv:hep-th/0502088];


\bibitem{casalderrey2010cherenkov} J. Casalderrey-Solana, D. Fern´andez, and D. Mateos,  \textit{ Cherenkov mesons as in-medium quark energy loss,  }    JHEP  {\bf 2010 },1 (2010)[arXiv:hep-th/1009.5937];



\bibitem{mateos2007thermodynamics} D. Mateos, R. C. Myers, and R. M. Thomson,  \textit{ Thermodynamics of the brane,  }   JHEP  {\bf 2007 },067 (2007)[arXiv:hep-th/0701132];


\bibitem{gwak2012holographic} B. Gwak, M. Kim, B.-H. Lee, Y. Seo, and S.-J. Sin,  \textit{ Holographic D instanton liquid and chiral transition,  }   Phys. Rev. D  {\bf 86 },026010 (2012)[arXiv:hep-th/1203.4883];


\bibitem{myers2007holographic}R. C. Myers, A. O. Starinets, and R. M. Thomson,  \textit{ Holographic spectral functions and diffusion constants for fundamental matter,  }   JHEP  {\bf 2007 },091 (2007)[arXiv:hep-th/0706.0162];



\bibitem{ejaz2008limiting} Q. J. Ejaz, T. Faulkner, H. Liu, K. Rajagopal, and U. A. Wiedemann,  \textit{ A limiting velocity for quarkonium propagation in a strongly coupled plasma via AdS/CFT,  }    JHEP  {\bf 2008 },089 (2008)[arXiv:hep-th/0712.0590];


\bibitem{casalderrey2010new}J. Casalderrey-Solana, D. Fern´andez, and D. Mateos,  \textit{ New Mechanism for Quark Energy Loss,  }    Phys. Rev. Lett.  {\bf 104 },172301 (2010)[arXiv:hep-ph/0912.3717];





\end{thebibliography}
  \end{document}